\documentclass[aip,apl,twocolumn,reprint,superscriptaddress, longbibliography]{revtex4-1}
\usepackage{graphicx}
\usepackage{color}
\usepackage{url}
\usepackage{amsmath}

\begin{document}
\title{Size-Dependent Grain Boundary Scattering in Topological Semimetals}

\author{Nicholas A. Lanzillo$^{a}$}
\affiliation{IBM Research, 257 Fuller Road, Albany, NY 12203, USA}
\thanks{Corresponding Author (nalanzil@us.ibm.com)} 

\author{Utkarsh Bajpai}
\affiliation{IBM Research, 257 Fuller Road, Albany, NY 12203, USA}

\author{Ion Garate}
\affiliation{Département de physique, Institut quantique and Regroupement Québécois sur les Matériaux de Pointe, Université de Sherbrooke, Sherbrooke, Québec J1K 2R1, Canada}

\author{Ching-Tzu Chen$^{b}$}
\affiliation{IBM T.J. Watson Research Center, 1101 Kitchawan Road, Yorktown Heights, NY 10598, USA}
\thanks{Second Corresponding Author (cchen3@us.ibm.com)}
\date{\today}

\begin{abstract}

We assess the viability of topological semimetals for application in advanced interconnect technology, where conductor size is on the order of a few nanometers and grain boundaries are expected to be prevalent. We investigate the electron transport properties and grain boundary scattering in thin films of the topological semimetals CoSi and CoGe using first-principles calculations combined with the Non-Equilibrium Green's Function (NEGF) technique. Unlike conventional interconnect metals like Cu and Al, we find that CoSi and CoGe conduct primarily through topologically-protected surface states in thin film structures even in the presence of grain boundaries. The area-normalized resistance decreases with decreasing film thickness for CoSi and CoGe thin films both with and without grain boundaries; a trend opposite to that of the conventional metals Cu and Al. The surface-dominated transport mechanisms in thin films of topological semimetals with grain boundaries demonstrates a new paradigm of the classical resistivity size-effect, and suggests that these materials may be promising candidates for applications as nano-interconnects where high electrical resistivity acts as a major bottleneck limiting semiconductor device performance. 
\end{abstract}

%\keywords{}
\maketitle

%%%%%%%%%%%%%%%%%%%%%%%%%%%%%%%%%%%%%%%%%%%%%%%%%%%%%%%%%%%%%%%%%%%%%%%%

\section{Introduction}
Interconnects in advanced technology nodes are subject to ever-increasing electrical resistivity as physical dimensions continue to aggressively shrink. This so-called resistivity size-effect is driven in large part by grain boundary scattering and surface scattering as interconnect dimensions shrink below the bulk electron mean free path\cite{fuchs1938conductivity,sondheimer1952mean,mayadas1970electrical,josell2009size}. High interconnect resistance acts as a bottleneck in achieving favorable semiconductor device performance and has motivated the search for alternate metals in next-generation logic technologies\cite{lanzillo2020iitc,nogami2017comparison,sankaran2015exploring,lanzillo2017abinitio}. While elemental metals such as Co, Ru, Mo and Pt have received considerable attention as potential replacements for state-of-the-art Cu interconnects, topological semimetals have recently been identified as promising candidates for interconnect applications due to their unusual scaling properties\cite{chen2020topological}. 
	
Topological semimetals are a class of matter which possess non-trivial bulk band structure topology, yielding disorder-tolerant surface states passing through the band crossing points\cite{armitage2018weyl,jia2016weyl,hasan2021weyl,burkov2011topological}. Unlike conventional metals, in topological semimetals the conduction and valence bands only touch at discrete nodal points (called \textit{Weyl nodes} when they are two-fold degenerate) or along one-dimensional lines in the Brillouin Zone. The band crossing points in topological semimetals appear in pairs with opposite chirality that are connected to surface states called \textit{Fermi Arcs}, which are a defining feature of this class of material\cite{jia2016weyl}. 

The surface states of topological semimetals can contribute to electron conduction with the same order-of-magnitude as bulk states\cite{breitkreiz2019large}, while being robust against small perturbations such as local defects, impurities and minor structural distortions\cite{chen2020topological}. In thin film samples of certain topological semimetals like CoSi, the surface conduction dominates over the bulk conduction, resulting in a different paradigm of resistivity scaling in sharp contrast to conventional metals\cite{chen2020topological}. 

In conventional metals, electron scattering increases as thin film thickness decreases due to the increased contributions of surface scattering and grain boundaries, giving rise to an increase in resistivity. In topological semimetals, e.g. CoSi and CoGe, however, we find that the resistivity decreases with decreasing film thickness, even in the presence of grain boundaries, owing to their surface-dominated transport channels. This trend, which is opposite to the well-known resistivity size-effect of conventional metals, makes topological semimetals of particular interest for technological applications where conductors with critically small dimensions are faced with ever-increasing electrical resistivity\cite{fuchs1938conductivity,sondheimer1952mean,josell2009size,smith2019evaluation}. For this reason, topological semimetals such as CoSi and NbAs are being explored as potential replacements for conventional metals like Cu in advanced interconnect technology\cite{zhang2019ultrahigh,chen2020topological}. 

In this paper, we evaluate the electron scattering characteristics at grain boundaries in thin films of two representative topological semimetals (CoSi and CoGe, both of which are prototypical multifold fermion semimetals\cite{tang2017multiple,chang2017unconventional,rao2019observation,sanchez2019topological,schroter2019chiral}) relative to two conventional interconnect metals (Cu and Al) whose grain boundary scattering is already well-understood from a theoretical point of view\cite{cesar2014calculated,zhou2018first,cesar2016reducing}. We find that total electron transmission decreases with decreasing film thicknesses in conventional metals as well as topological semimetals whether or not a grain boundary is present, but with different rates for topological vs. conventional metals. As a result, in the cases of CoSi and CoGe, we find that grain boundary specific resistivity ($\gamma_{GB}$) actually decreases with decreasing film thickness, whereas the opposite trend is observed for Cu and Al. This is attributed to the topologically-protected surface states in CoSi and CoGe, which are not as susceptible to bulk-like grain boundary scattering mechanism observed in both Cu and Al. These results suggest that topological semimetals may be promising candidates for next-generation interconnects with critically small feature size. 

This paper is organized as follows: In section II we discuss the computational methods employed in our study. In section III(A), we discuss grain boundary scattering in the bulk for conventional metals and topological semimetals. In section III(B), we then discuss grain boundary scattering in thin films of conventional metals and topological semimetals. Finally, we have concluding remarks in section IV. 

\section{Methods} 
We performed simulations based on density functional theory (DFT) as implemented in the Synopsys QuantumATK software package\cite{quantumatk}. The generalized gradient approximation (GGA) was employed for the exchange-correlation functional\cite{perdew1997generalized} and we used a cutoff energy of 75.0 Hartree for all elements as well as the double-zeta-polarized basis set, which has been demonstrated to give accurate results for the metallic elements considered here\cite{lanzillo2018electron,lanzillo2019electron,hegde2016lower,lanzillo2017electronic,jones2015electron}. Spin-orbit interactions are small in these materials, so they have been neglected in our calculations. 

We consider both Cu and Al in the face-centered-cubic crystal structure with equilibrium lattice constants of 3.61\AA{} and 4.05\AA{}, respectively. We take the simple cubic crystal structure of both CoSi and CoGe with lattice constants of 4.443\AA{} and 4.653\AA{}, respectively (see Fig. 1 below.) For ease of comparison, we consider twin (100)/(100) grain boundaries in all four structures, consisting of two (100) surfaces with normal vectors anti-parallel to one another (shown later in Section II.) Relaxed grain boundary structures are obtained by fixing all atoms except for the 4 atomic layers closest to the grain boundary interface. The fixed atoms are allowed to move as a rigid unit but cannot relax individually. Lastly, the geometries of all structures were relaxed until the forces acting on the ions were less than 0.05 eV/\AA{} prior to transport calculations. 

We consider unstrained crystals. Recent electronic structure calculations\cite{bose2021strain} show that in the presence of biaxial strain, the multifold fermions of CoSi split into multiple Weyl fermions. The Fermi arcs located on the (001) surface are found to be robust under strain, at least up to 2\%. Based on this finding, we anticipate that strain will have a quantitative, but not qualitative, impact on our results. 

Quantum transport calculations were carried out using the Non-Equilibrium Green's Function (NEGF) approach\cite{hang2008quantum}. Structures used for transport consisted of semi-infinite left/right electrode regions and a central scattering region. For bulk transport calculations, we used k-point sampling of 11x11x301 for Cu and Al and sampling of 5x5x301 for CoSi and CoGe. For thin-film calculations we used k-point sampling of 1x7x301 for all structures. We checked that increasing the number of $k$-points affects the calculated values of transmission by less than 2\%. 

In the NEGF formalism employed in this work, the electronic conductance $G$ (per spin channel) is calculated using the linear-response expression for the electronic current $I = G V$, where $V$ is the small externally applied bias-voltage. In such a scenario
\begin{equation}
G = \frac{1}{R} = \frac{e^2}{h} T(E_F),
\end{equation} 
where $T(E_F) = \sum_{i} T_{i}(E_F)$ is the total electron transmission obtained as a sum over $T_i(E_F)$ that represents the electron transmission for a transport channel $i$; $E_F$ is the Fermi-Energy of the system; and $R$ is the corresponding resistance. For the thin films considered in our work, the resistance $R$ helps us to evaluate the so-called grain boundary specific resistivity $\gamma_\mathrm{GB}$ defined using the well-established convention\cite{cesar2014calculated,cesar2016reducing,lanzillo2017abinitio,lanzillo2018defect}.
\begin{equation}\label{eq:gamma}
\gamma_\mathrm{GB} = A_\text{thin-film}(R_\mathrm{GB} - R_\mathrm{pristine}),
\end{equation}
where $A_\text{thin-film}$ is the area of cross-section of the thin film, $R_\mathrm{GB}$ is its resistance in the presence of a grain boundary, and $R_\mathrm{pristine}$ is the corresponding resistance for a pristine sample where the grain boundary is absent. Equation~\eqref{eq:gamma} allows one to isolate the resistance contribution that originates solely due to the grain boundary by subtracting out the Sharvin resistance~\cite{gall2016electron} of the pristine thin films where grain boundary is absent.

\section{Results and Discussion}
We first characterize bulk grain boundary scattering in both conventional metals (Cu, Al) and topological semi-metals (CoSi, CoGe) at a representative grain boundary. Since Cu, Al, CoSi and CoGe all share a cubic crystal structure, we construct the (100)/(100) grain boundary for all of them, followed by a characterization of electron transport across the (100)/(100) grain boundary. We found the (100) surface of CoSi has the lowest surface energy (12.8 eV/nm$^2$) relative to other possible surface terminations, which motivates the choice of the (100)/(100) grain boundary, and the same grain boundary was chosen for the conventional metals (which have a face-centered-cubic lattice) for ease of comparison. 

\subsection{Electron Scattering at Grain Boundaries in Bulk Metals} 
We begin by characterizing the bulk electronic structure of our two representative topological semimetals: CoSi and CoGe. Both semimetals have a simple cubic crystal structure and the corresponding Brillouin Zone, as depicted in Fig. 1(a) and (b), respectively. The electronic band structures of CoSi and CoGe are shown in Fig. 1 (c) and (d), respectively. 
\begin{figure}[h!]
	\centering%
	\begin{center}
		\includegraphics[scale=0.35,clip,trim = 0mm 0mm 0mm 0mm]{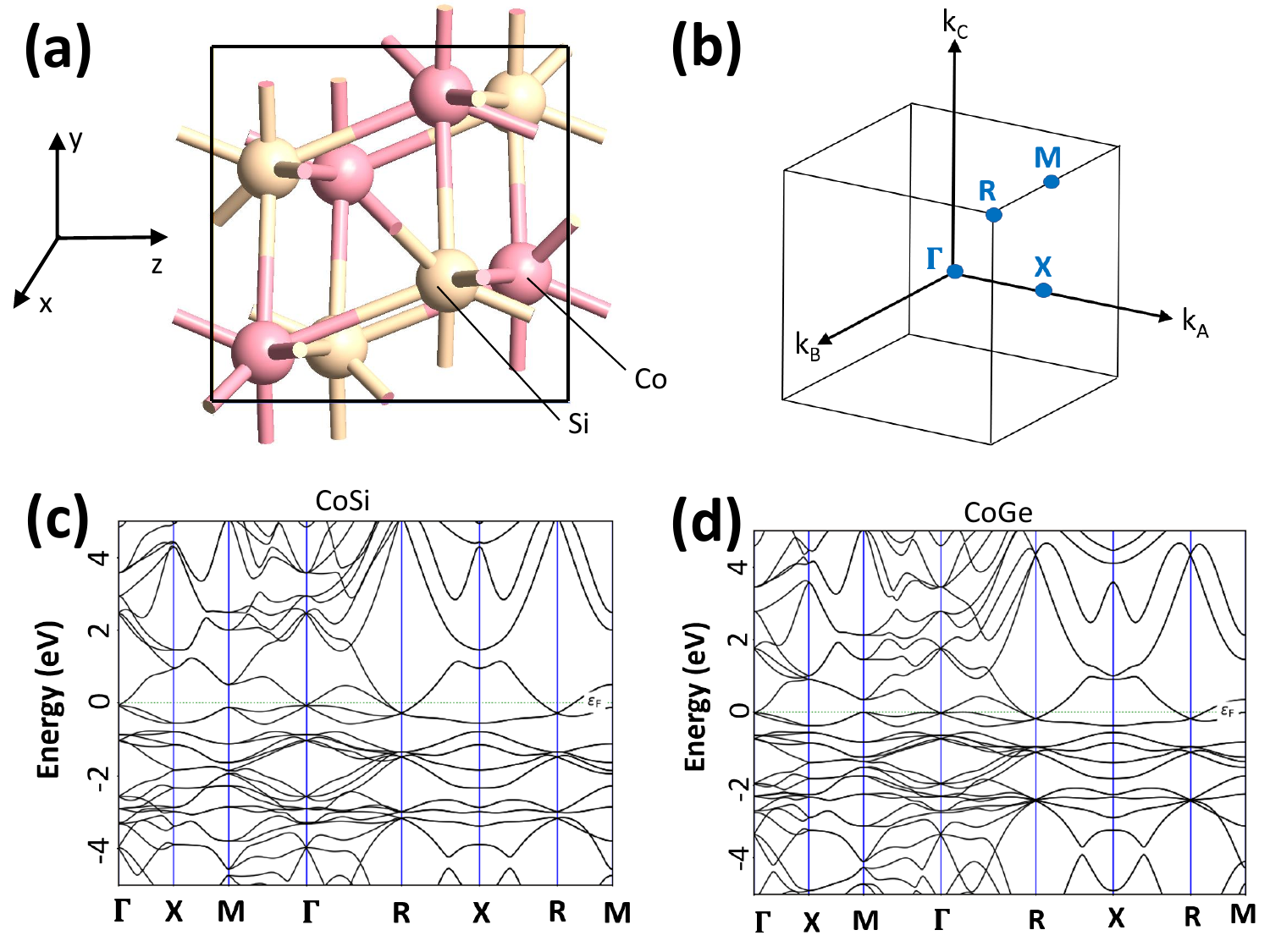}
	\end{center}
	\caption {(a) Unit cell depicting the simple cubic crystal structure of CoSi and CoGe. (b) Brillouin Zone depicting points of high symmetry. (c) Electronic band structure of bulk CoSi and (d) CoGe.}
\end{figure}
We see that both semimetals have bands crossing near the Fermi Energy. The band crossing at the zone center ($\Gamma$-point) and at the zone corner (R-point) form a pair of topological band crossings. The associated Chern numbers of $\pm 2$ guarantee that there are two surface bands per spin and per surface connecting between the nodes, providing additional conducting channels in conjunction to channels from bulk bands. 

For the case of Cu as a representative conventional metal, an atomic-scale representation of the (100)/(100) grain boundary is shown in Fig. 2(a) with the electrodes for NEGF calculations shown on either side. We consider electron transmission across this interface ($T_{GB}$) as well as transport along the (100) direction in a pristine sample ($T_{pristine}$) where the grain boundary is absent. Values of transmission are converted to resistance using eq. (1) such that $G = \frac{2e^2}{h} T_{total} = 1/R$ where $T_{total} = \sum_{i} T_{i}$. The k-resolved electron transmission for both the pristine (100) case and the (100)/(100) grain boundary in bulk Cu are shown in Fig. 2(b) and (c), respectively. 
\begin{figure}[h!]
	\centering%
	\begin{center}
		\includegraphics[scale=0.35,clip,trim = 0mm 0mm 0mm 0mm]{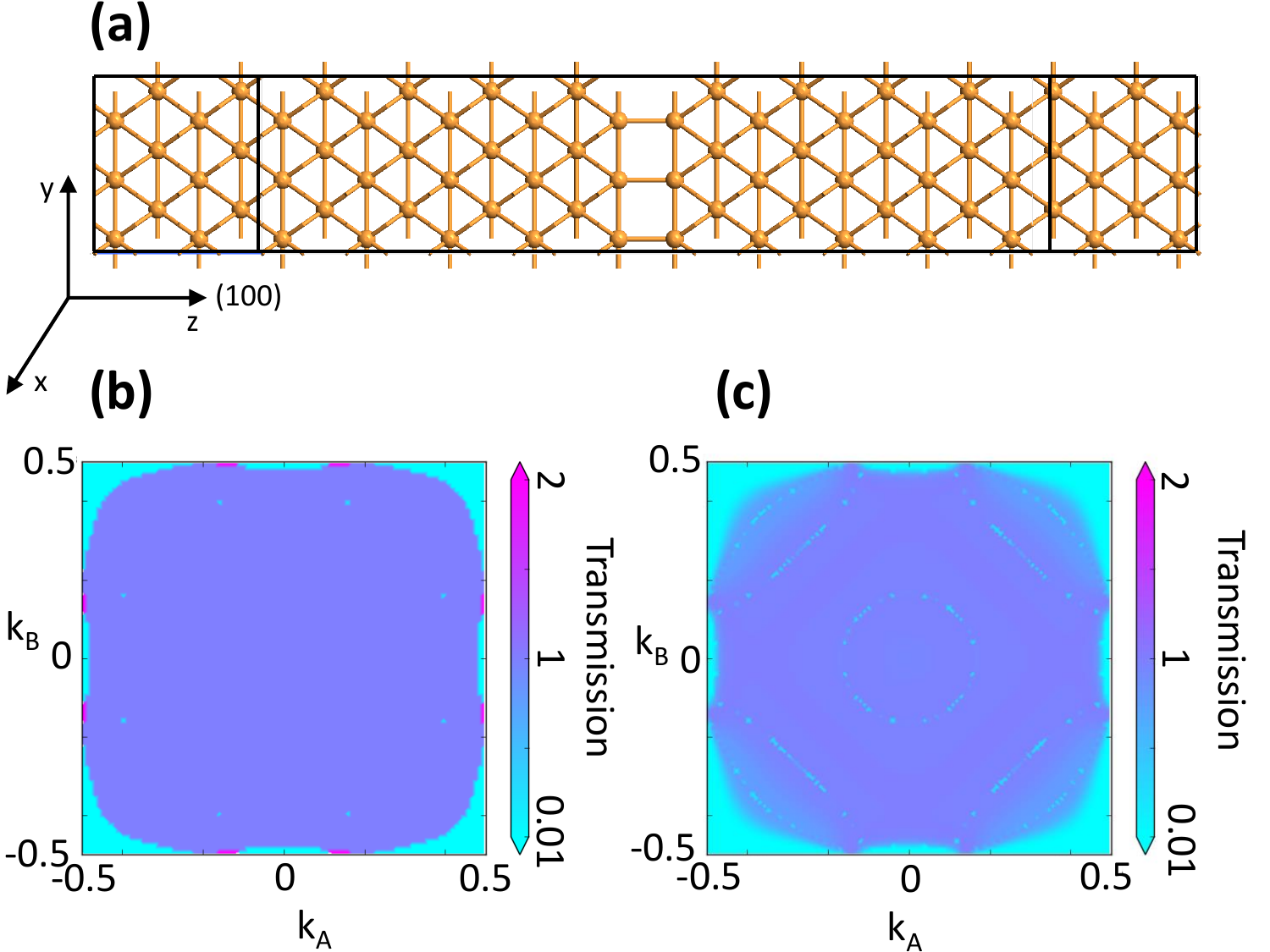}
	\end{center}
	\caption {Impact of grain boundary on electron scattering in Cu. (a) Depiction of a (100)/(100) grain boundary in bulk Cu. (b) Momentum-resolved transmission spectra per spin for electron transport along bulk Cu(100) without a grain boundary. (c) Momentum-resolved transmission spectra per spin for electron transport across a bulk Cu (100)/(100) grain boundary. In (b) and (c), $k_{A}$ and $k_{B}$ denote the momenta in the plane perpendicular to the transport direction. Each k-point contributes only one channel of transmission (per spin) in the forward direction because only one spin-degenerate band crosses the Fermi level.}
\end{figure}
For electron transport along (100) in pristine Cu, we see essentially uniform electron transmission with $T(k_{A},k_{B})=1$ throughout the Brillouin Zone, as shown in Fig. 2(b), which is consistent with the nearly isotropic Fermi surface in Cu. However, when the (100)/(100) grain boundary is introduced [as shown in Fig. 2(c)] we see localized regions in Brillouin Zone with lower transmission, but most of the other regions are undisturbed. 

Next, we turn to the electron transmission across the (100)/(100) grain boundary in bulk CoSi. An atomic-scale representation of the grain boundary as well as the electrode regions for NEGF calculations is shown in Fig. 3(a).
\begin{figure}[h!]
	\centering%
	\begin{center}
		\includegraphics[scale=0.35,clip,trim = 0mm 0mm 0mm 0mm]{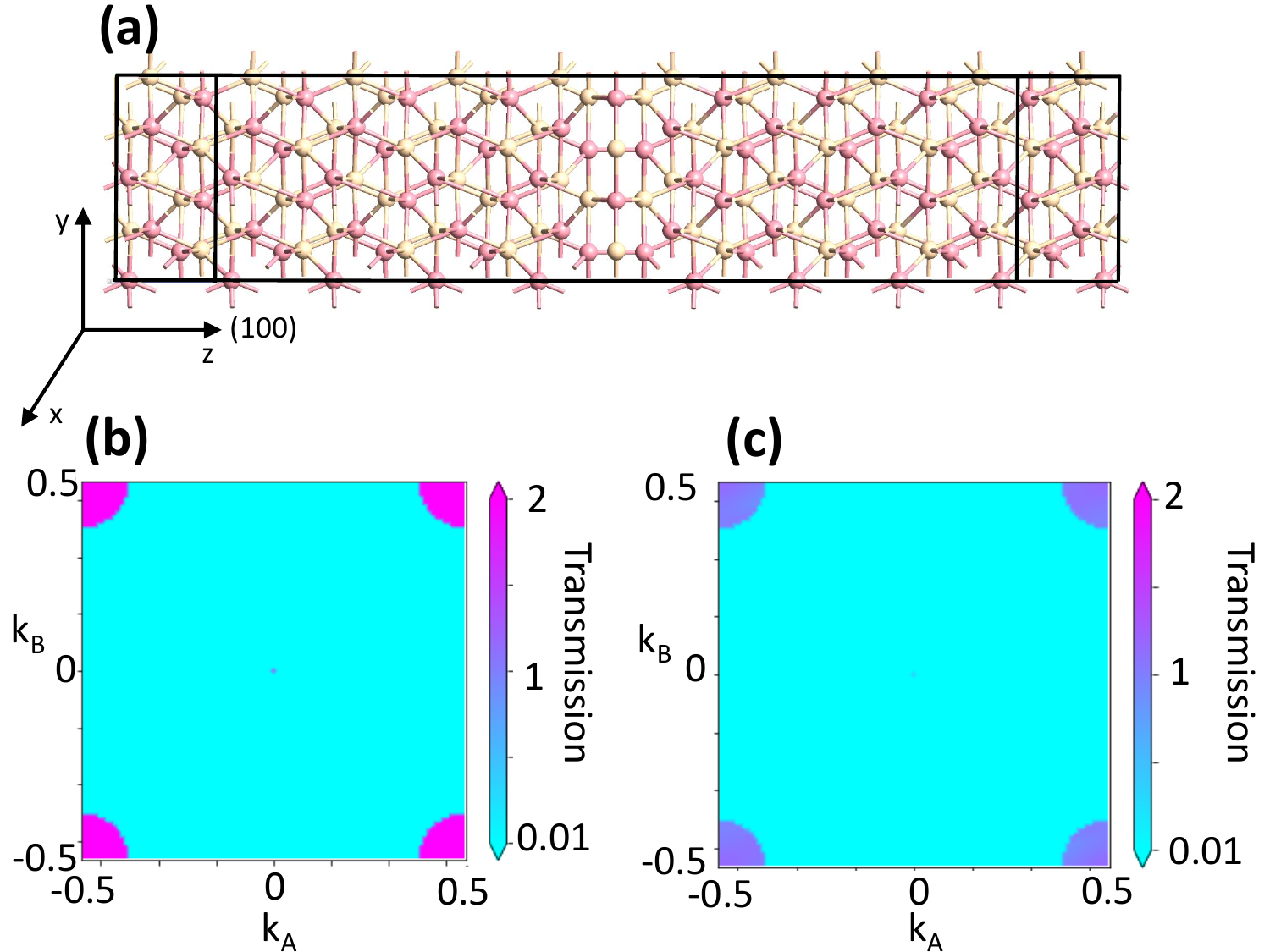}
	\end{center}
	\caption {Impact of grain boundary on electron scattering in CoSi (a) Depiction of a (100)/(100) grain boundary in bulk CoSi. (b) Momentum-resolved transmission spectra for electron transport along bulk CoSi(100) without a grain boundary. (c) Momentum-resolved transmission spectra for electron transport across a bulk CoSi (100)/(100) grain boundary. Here, $k_{A}$ and $k_{B}$ denote the momenta in the plane perpendicular to the transport direction. In contrast to Cu, each k-point near the zone corners contributes two channels of transmission (per spin) in the forward direction because two spin-degenerate bands cross the Fermi level.}
\end{figure}
In stark contrast to the case of Cu in Fig. 2, we see that the k-resolved electron transmission for transport along the (100) direction in bulk CoSi is essentially zero everywhere in the Brillouin Zone except at the pockets in  the corners where the bulk bands cross the Fermi level [as seen in Fig. 1(c)]. This is because in contrast to normal metals, the bulk bands in CoSi cross the Fermi level only near the zone center and the zone corners, with a negligible density of states (DOS) near the $\Gamma$-point [as seen in Fig. 1(c)], resulting in a significantly smaller number of bulk states available for transport. On the other hand, for transport across the (100)/(100) grain boundary in bulk CoSi, we see that the electron transmission at the pockets in the corners of the Brillouin Zone is sustained but its magnitude is reduced a value of 2.0 to 1.0 [Figs. 3(b) and 3(c)], consistent with a roughly 50\% reduction of average transmission in CoSi when a grain boundary is introduced (Table I.) The properties related to bulk grain boundary scattering for our conventional metals and topological semimetals are listed in Table I. 
\begin{table}
	\begin{tabular}{|c|cc|cc|cc|cc|}
		\hline 
		Metal &  Cu      &       & Al      &      & CoSi     &       & CoGe   &  \\
		& P   &    GB  & P  & GB    & P & GB     & P &  GB\\  
		\hline  
		Area ($A$)   &   6.53          & 6.53  & 8.18           & 8.18 &  19.69          & 19.69  & 21.62 & 21.62        \\ 
		$T$  &   0.92          & 0.80  &  1.63          & 1.01 &  0.09          & 0.05  & 0.21  & 0.06  \\ 
		$RA$ &  9.16 & 10.53 & 6.47 & 10.45 & 282.37 & 508.26 & 132.87 & 465.07 \\
		$\gamma_{GB}$ &  & 1.37 & & 3.97 & & 225.89 & & 332.19 \\ 
		\hline 
	\end{tabular}
	\caption{A summary of the transmission calculations for bulk Cu, Al, CoSi and CoGe. P=pristine and GB=grain boundary. $A$ refers to the cross-sectional area of the supercell and is in units of $\AA{}^2$, transmission is dimensionless, and both $RA$ and $\gamma_{GB}$ are in units of $1 \times 10^{-12} \Omega cm^2$. Average transmission is defined as the transmission per spin averaged over the $k$-points in the projected 2D Brillouin Zone perpendicular to the transport direction.} 
\end{table} 

\subsection{Electron Scattering at Grain Boundaries in Thin Films}
Next, we investigate electron scattering in thin film structures both with and without grain boundaries. These films are intended to be representative of ultra-scaled interconnects used in advanced semiconductor technologies, where conductor feature size is on the order of a few nanometers and non-idealities such as grain boundaries are expected to be prevalent. For the cases of Cu and Al, we constructed thin film unit cells with film thickness varying between 3 atomic layers (less than 1 nm) and 15 atomic layers (more than 4nm) and the electron transport direction is oriented along (100) direction in all cases (the $z$-axis is parallel to (100) in Figs. 4 and 5.) For a given thin film thickness, we consider the pristine structures as well as the structures with a (100)/(100) grain boundary present. Atomic-scale representations of the thin film structures for the case of Cu are shown in Fig. 4. The supercells for Al are identical except for the lattice constant. 
\begin{figure}[h!]
	\centering%
	\begin{center}
		\includegraphics[scale=0.35,clip,trim = 0mm 0mm 0mm 0mm]{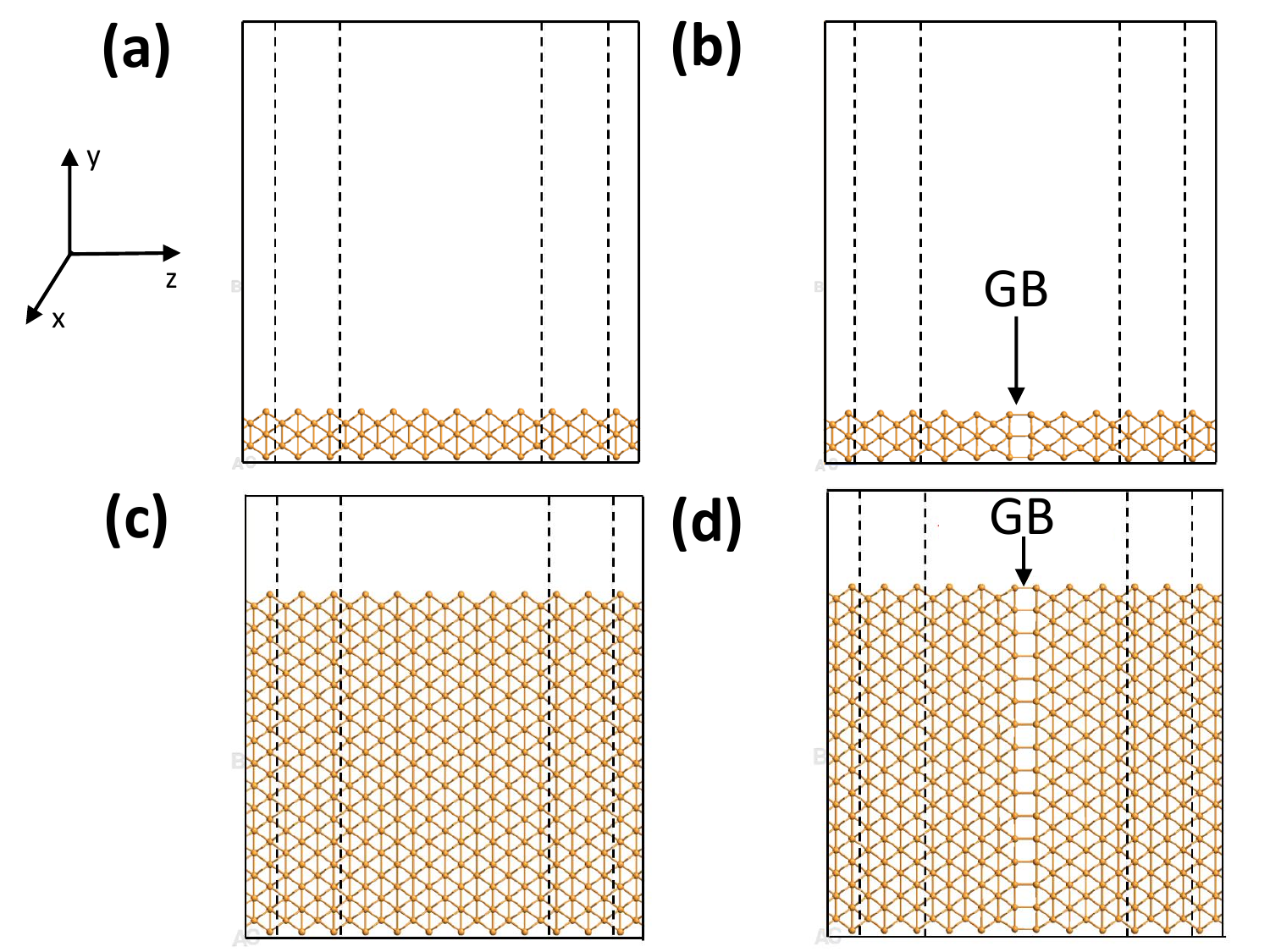}
	\end{center}
	\caption {Cu thin film structures used for transport calculations, ranging from the thinnest, 3-atomic layer thick structures without (a) and with (b) a (100)/(100) grain boundary, as well as the thickest, 15-atomic layer thick structures without (c) and with (d) a (100)/(100) grain boundary. The dotted lines denote the semi-infinite electrode regions. The transport direction (100) is oriented along the $z$-axis.}
\end{figure}
All four structures are cubic, which implies that the surface termination of the thin films is such that the normal vector also points along the (100) direction. We found no appreciable geometric relaxation in any of the Cu or Al thin film structures other than the few atomic layers immediately surrounding the grain boundary. In a similar fashion, we constructed supercells for CoSi and CoGe thin films ranging from 3 atomic layers to 15 atomic layers. Atomic-scale images of the supercells for the case of CoSi are shown in Fig. 5. 
\begin{figure}[h!]
	\centering%
	\begin{center}
		\includegraphics[scale=0.35,clip,trim = 0mm 0mm 0mm 0mm]{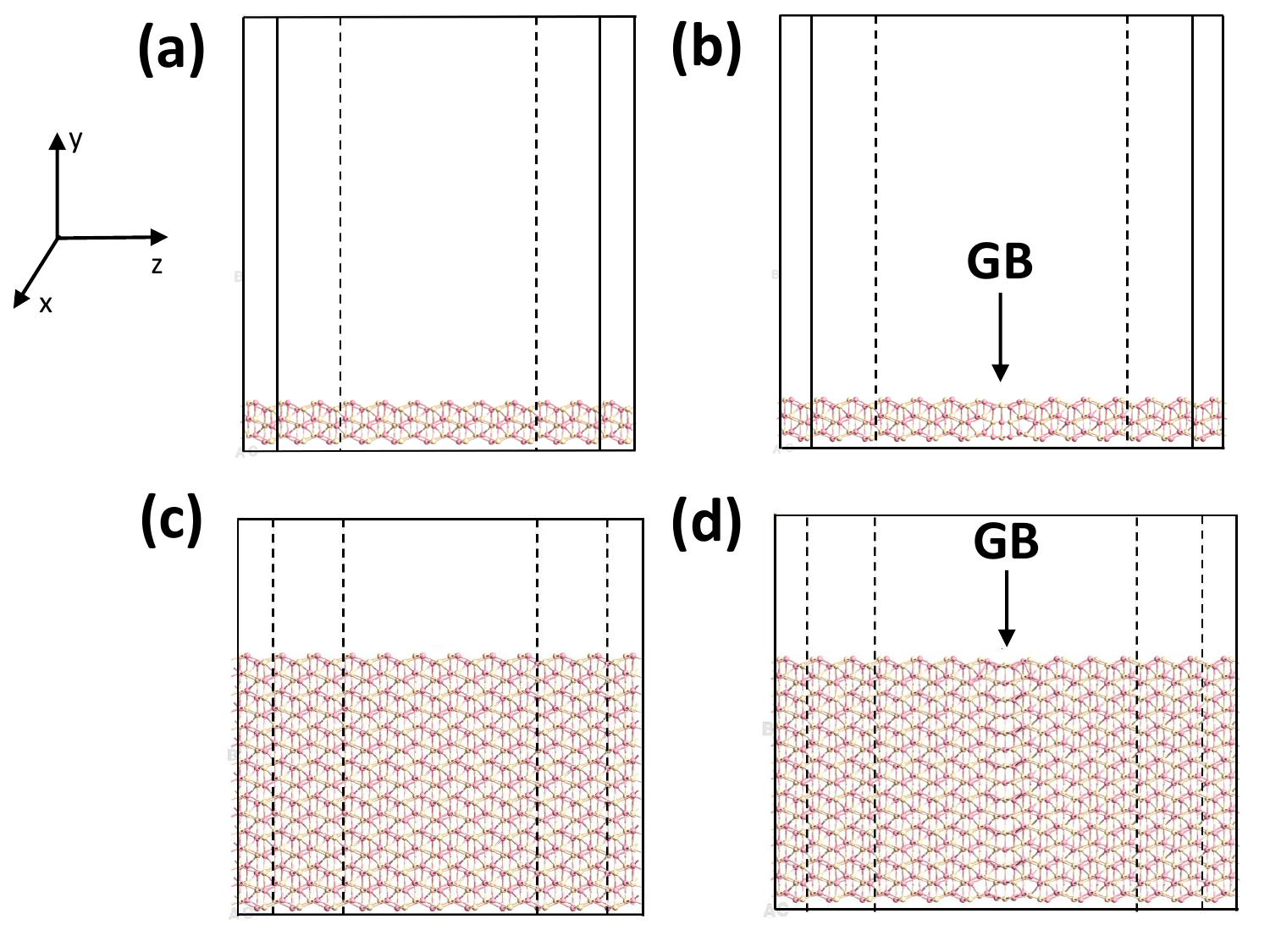}
	\end{center}
	\caption {CoSi thin film structures used for transport calculations, ranging from the thinnest, 3-atomic layer thick structures without (a) and with (b) a (100)/(100) grain boundary, as well as the thickest, 15-atomic layer thick structures without (c) and with (d) a (100)/(100) grain boundary. The dotted lines denote the semi-infinite electrode regions.}
\end{figure}
Similar to the cases of Cu and Al, we only find minor geometric relaxation in the immediate vicinity of the grain boundary, and the atoms remain in their bulk-like configuration. 

For each thin film, we calculate the electron transmission using the NEGF formalism described in the Methods section. The calculated values of electron transmission for each thin film structure both with and without a grain boundary are plotted in Fig. 6. 
\begin{figure}[h!]
	\centering%
	\begin{center}
		\includegraphics[scale=0.35,clip,trim = 0mm 0mm 0mm 0mm]{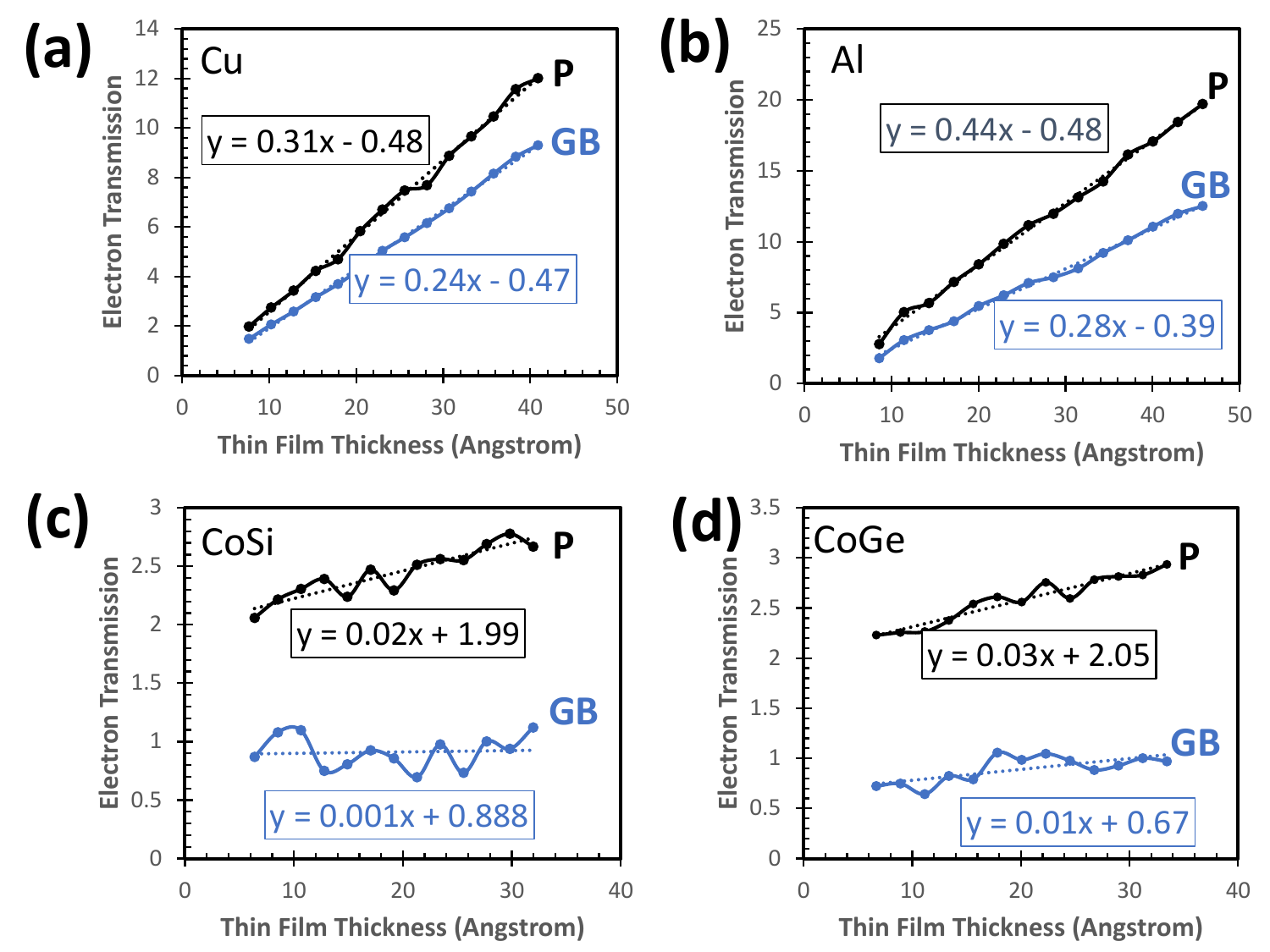}
	\end{center}
	\caption {Electron transmission as a function of thin film thickness for (a) Cu (b) Al (c) CoSi and (d) CoGe. The black lines are for pristine thin films and the blue lines are for thin films with (100)/(100) grain boundaries. The equations of the best fit straight lines are included for each case.}
\end{figure}
We see that for our conventional metals [Figs. 6(a) and (b)] as well as topological semimetals [Figs. 6(c) and (d)], the electron transmission increases linearly with thin film thickness. For the case of the conventional metals Cu and Al, this is consistent with previous reports in the literature in which the number of conducting channels has been found to increase linearly with conductor cross-sectional area, indicative of the increasing number of quantum-well (bulk) states in the thin films\cite{zhou2008resistance,simbeck2012aluminum,zhou2010abinitio,hegde2016lower}. Furthermore, we perform a straight line fit ($y=mx+b$) to the electron transmission curves in Fig. 6, where the slope of the straight lines for CoSi and CoGe are an order-of-magnitude smaller than the slopes for Cu and Al, indicating that the electron transmission grows more slowly as a function of film thickness. We attribute this to the comparatively low bulk electronic DOS in CoSi and CoGe at the Fermi level [as discussed in Figs. 3(b) and (c).] 

Typically, as the film thickness increases, the bulk-like contribution plays a larger role, but this impact is muted in the topological semimetals relative to the conventional metals. In contrast, we see that while the $y$-intercepts (representing the transmission at zero-film thickness) for the two conventional metals is nearly zero, it almost exactly approaches a value of two for the pristine topological semimetals CoSi and CoGe [Fig. 6(c) and 6(d).] This indicates that there are two channels of surface states (per spin) contributing to the electron transmission in pristine CoSi and CoGe slabs. The surface states are the only remaining contribution to transport in the limit of a very thin film; as long as the film thickness is larger than the total penetration depth of the surface states of $\approx$5 atomic layers (see Fig. 7 for the local density of states of CoSi at the Fermi level as a function of depth from the surface), contributions of the surface states to transmission remain largely constant. Furthermore, in the thickness range under study, the surface-state contribution to the total transmission dominates over the bulk-state contribution in CoSi and CoGe thin films, which reflects that the Fermi arcs of the topological surface states span a much larger region in k-space than the Fermi surfaces of the bulk states\cite{chen2020topological}. 

For our conventional metals as well as topological semimetals, the slopes of the curves representing electron transmission across a (100)/(100) grain boundary are less than the corresponding curves for pristine samples. This is due to grain boundary scattering. There is again an order-of-magnitude difference in the slope between grain boundary transmission curves for conventional metals relative to CoSi and CoGe, reflecting the order-of-magnitude difference in their bulk DOS. Furthermore, in CoSi and CoGe, the $y$-intercepts of transmission per spin also reduce, indicative of grain boundary scattering of the topological surface state electrons. 

To better illustrate the contrasting bulk-dominated vs. surface-dominated transport in conventional metals v. topological semimetals, we plot the local electronic density of states (LDOS) for 10-atomic-layer-thick thin films of both Cu and CoSi with grain boundaries in Fig. 7. 
\begin{figure}[h!]
	\centering%
	\begin{center}
		\includegraphics[scale=0.27,clip,trim = 0mm 0mm 0mm 0mm]{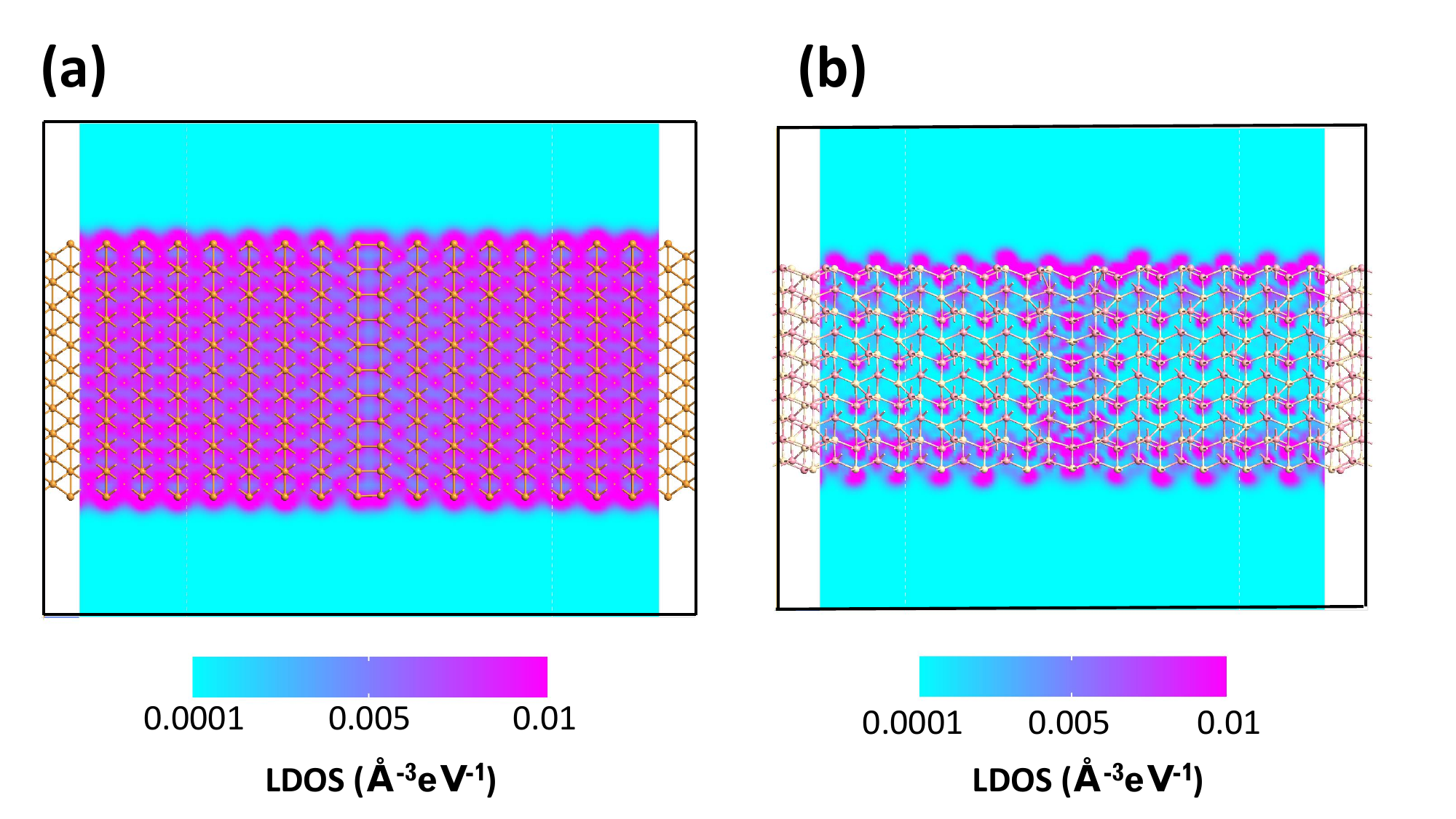}
	\end{center}
	\caption {Two-dimensional projection of the local density of states (LDOS) at the Fermi level for 10-atomic-layer thick thin films of (a) Cu and (b) CoSi.}
\end{figure}
It is immediately clear upon inspection of Fig. 7 that Cu has a relatively uniform LDOS at the Fermi level throughout the interior of the thin film with a slight increase toward the surfaces, while the CoSi thin film has a significantly higher LDOS at the surfaces compared to the interior of the thin film. The CoSi thin film also shows increased LDOS in the region immediately surrounding the grain boundary, providing pathways for backscattering of the topological surface states, consistent with the reduced $y$-intercepts in Figs. 6(c) and 6(d). 

Next, we discuss the k-resolved transmission for thin films of Cu and CoSi as representative conventional metals and topological semimetals, respectively. The k-resolved transmission for thin films with thicknesses of 5, 10 and 15 atomic layers of both Cu and CoSi are shown in Fig. 8, where black lines correspond to pristine thin films and blue lines correspond to thin films with (100)/(100) grain boundaries present.  
\begin{figure}[h!]
	\centering%
	\begin{center}
		\includegraphics[scale=0.35,clip,trim = 0mm 0mm 0mm 0mm]{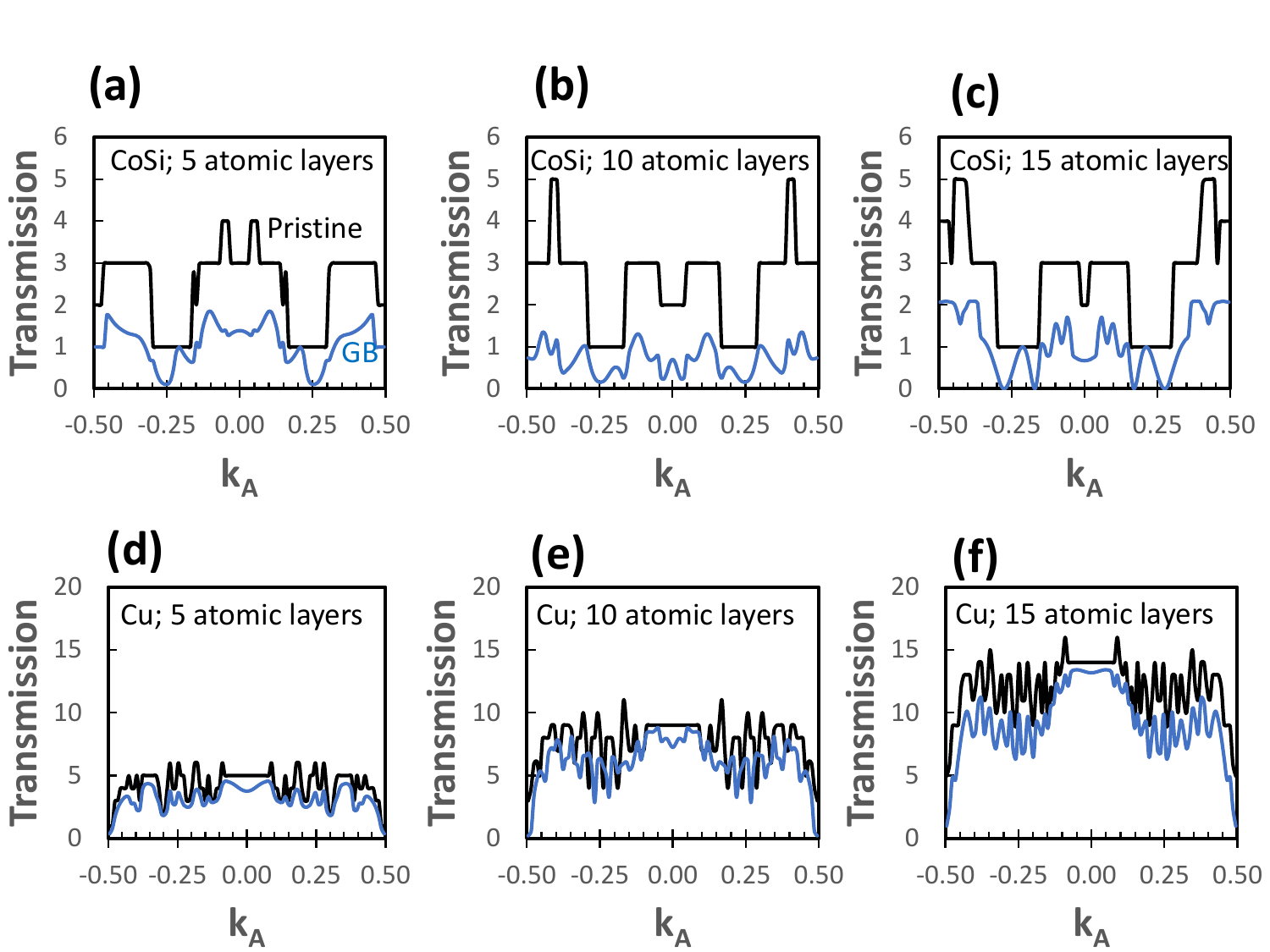}
	\end{center}
	\caption {K-resolved transmission spectra for Cu and CoSi thin film structure, ranging from 5 atomic layers thick, 10 atomic layers thick and 15 atomic layers thick. Black curves are for pristine thin films while blue curves are for thin films with (100)/(100) grain boundaries.}
\end{figure}
The $x$-axis is the Brillouin Zone monentum $k_{A}$, which is perpendicular to the transport direction and parallel to the film surface. For the case of CoSi [Figs. 8(a) through (c)] we observe only discrete values of electron transmission along $k_{A}$ in the pristine thin films, demonstrating the quantized nature of electron transport in a defect-free structure. We also observe that as the thin film thickness increases, the two localized peaks in Fig. 8(a) with the highest electron transmission gradually shift toward the boundaries of the Brillouin Zone, where $k_x = 0.5$. This trend originates from the increasing number of bulk quantum-well states near the zone corner with film thickness. Most of the bulk bands crossing the Fermi level emanate from the zone corner because the topological band crossing at the R-point [Fig. 1(c)] is approximately 150 meV below the Fermi level. 

In the limit of infinite film thickness, we expect these peaks to become localized near the Brillouin Zone corner, which is precisely what we observe in the bulk k-resolved transmission shown in Fig. 3(b), where we only see pockets of high transmission in the corners of the of the 2D Brillouin Zone. We also note that the transmission plateaus of $T=1$, $T=2$ and $T=3$ per spin for the pristine films do not change as the thin film thickness is modulated, which is because the transmission in this region originates from the surface state transmission. When a grain boundary is introduced in the thin films, the surface states are no longer topologically protected and can travel along the grain boundary to the opposite surface. This opens up backscattering channels and results in degraded electron transmission observed in the blue curves in Fig. 8(a)--(c). 

The k-resolved transmission of the Cu thin films shows different behavior than the CoSi thin films. While the values of transmission [indicated by the peaks and valleys in Fig. 8(c) through (e)] are still quantized at integer values of electron transmission, we see many more peaks and valleys, as well as significantly higher electronic density of states in Cu relative to CoSi due to the significantly higher DOS at the Fermi level. In addition, we see that the average magnitude of the electron transmission rapidly increases as the film becomes thicker, which is the result of more bulk-like channels becoming available as the thin film thickness increases. This is in stark contrast with the CoSi case, where the total transmission remains nearly unchanged with varying film thickness because electrons predominantly conduct through surface states, whereas bulk states only contribute through small pockets in the corners of the Brillouin Zone. 

Lastly, we quantify the impact of grain boundary scattering in thin film structures by calculating the specific resistivity of each grain boundary ($\gamma_{GB}$) as a function of film thickness. The calculation of $\gamma_{GB}$ is given in Eq. (2). The values of $\gamma_{GB}$ for Cu, Al, CoSi and CoGe thin films are plotted as a function of film thickness up to 5nm in Fig. 9. 
\begin{figure}[h!]
	\centering%
	\begin{center}
		\includegraphics[scale=0.35,clip,trim = 0mm 0mm 0mm 0mm]{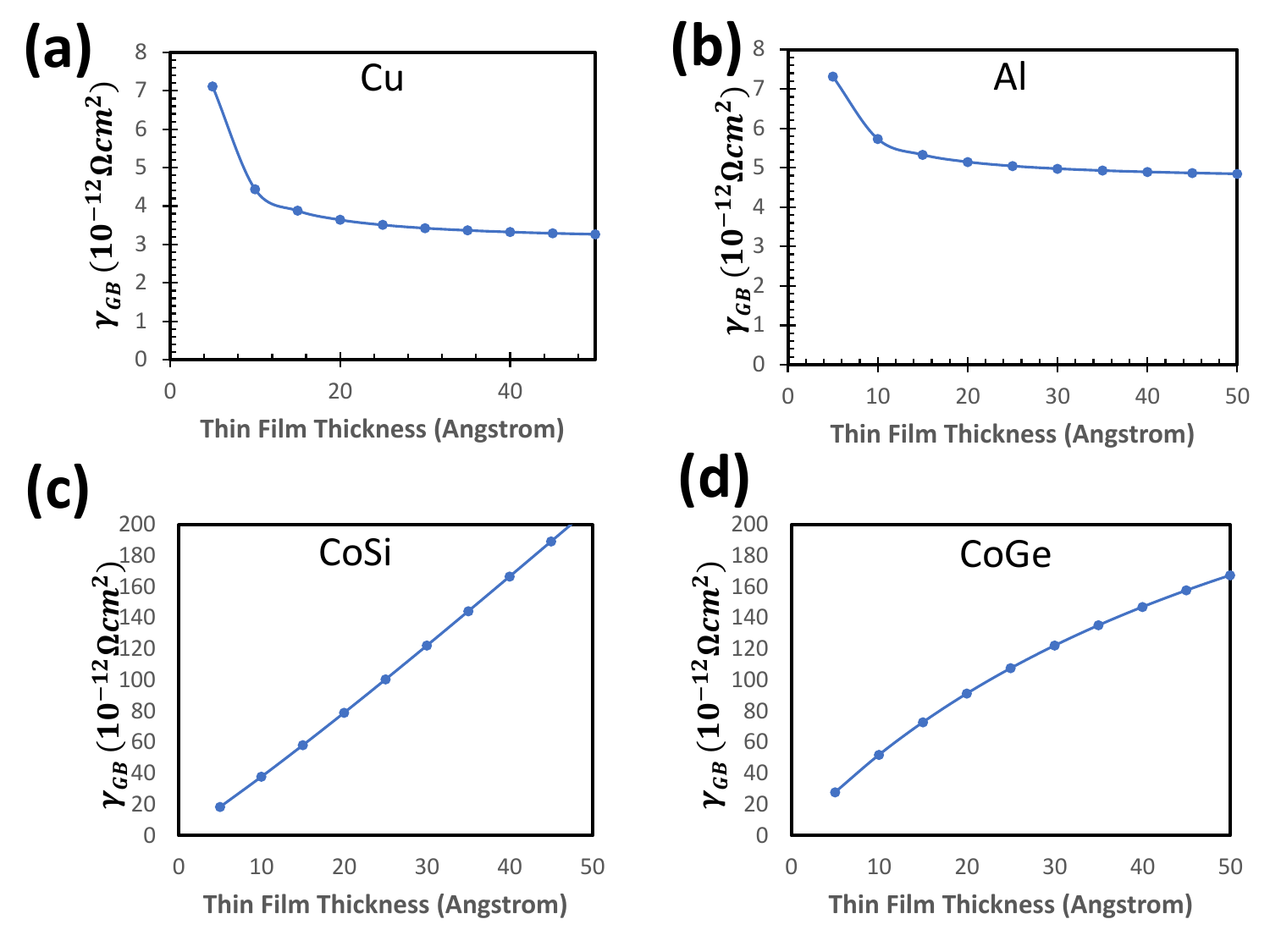}
	\end{center}
	\caption {Grain boundary specific resistivity as a function of thin film thickness for (a) Cu (b) Al (c) CoSi and (d) CoGe. Note that, with decreasing film thickness, $\gamma_{GB}$ increases in Cu and Al, while it decreases in CoSi and CoGe. Besides, the bulk limit of grain boundary specific resistivity is reached at a much smaller thickness in Cu and Al because their carrier densities are much higher and, thus, the Fermi wavelengths are much smaller than in CoSi and CoGe.}
\end{figure}
We see that the values of $\gamma_{GB}$ are essentially flat for the conventional metals Cu and Al until the film thickness becomes extremely small ($<2$nm), in which case the values start to increase. For the cases of CoSi and CoGe, on the other hand, we see that the values of $\gamma_{GB}$ decrease significantly as the film thickness decreases; behavior opposite to what is observed in the conventional metals. In bulk CoSi and CoGe, where transport is dominated by the bulk states, grain boundary scattering has a significant impact on electron transmission and on the bulk RA product. 
In thin films, however, the surface-to-bulk ratio gets larger and the topologically-protected surface states dominate the overall electron transport. The total transmission of the topological surface states across the grain boundary remains nearly constant regardless of the film thickness. Therefore, the total conductance per unit area increases with decreasing film thickness even in the presence of grain boundaries, giving rise to decreasing $\gamma_{GB}$ [cf. Figs. 9(c) and 9(d).]

\textcolor{black}{
While prior studies have shown that ultra-thin films of topological insulators and certain topological semimetals eventually self-gap as the result of quantum confinement and coupling/overlap between Dirac states on opposite surfaces\cite{chen2017topological,bian2012interfacial,zhang2012topological}, others have observed robust gapless surface states and remnant conductivity in the zero-thickness limit in topological semimetal Sb\cite{cairns2015observation}. Our simulation results for CoSi and CoGe indicate that these materials retain conducting surface states even in the ultra-thin limit. This can be seen in Fig. 6(c) and (d), where the non-zero y-intercept represents remnant conductivity in the zero-thickness limit, consistent with the decreased grain boundary specific resistivity in Fig. 9(c) and (d). In contrast, in the presence of an ultra-thin-film-induced band gap opening, we would expect the grain boundary specific resistivity curves in Fig. 9(c) and (d) to increase rather than decrease as thickness approaches zero. As a final confirmation, we have calculated the two-dimensional Fermi Surface for 4-atomic-layer thick CoSi, where we see that the surface states persist (Supplemental Information.) Since these conducting states originate from the topological surface states (not the bulk states), the number of the conducting channels does not extrapolate to zero in the zero-thickness limit.
}

While the grain boundaries studied in this work are simplified structures consisting of two anti-parallel (100) surfaces, recent studies have highlighted the impact of tilt on grain boundary specific resistivity, finding correlation between dislocation density, local strain and resistivity\cite{bishara2021understanding}. We anticipate that more complicated grain boundary structures in topological semimetals may have quantitatively different values of specific resistivity but are expected qualitatively to follow the same scaling trends identified in this work. 

In this work, we have only studied models without spin-orbit coupling (SOC) for CoSi and CoGe. When SOC is considered, Chern numbers of $\Gamma$ and $R$ become $\pm 4$\cite{chang2017unconventional}. The topological surface bands are split and their E-k dispersions vary slightly. Nevertheless, the total number of the Fermi arcs participating in transport remains the same: again, four arcs in total, two from [0,0] to [$\pi$, $\pi$] and two from [0,0] to [-$\pi$, -$\pi$]. Both the number of conducting channels and the traveling directions of electrons in the split surface bands remain the same as those of the unsplit bands. Therefore, qualitatively the trend of decreasing RA and decreasing grain-boundary specific resistivity with scaling persists in the presence of SOC> Experimentally, since the SOC-induced energy splitting is within the energy broadening of the CoSi samples, ARPES cannot resolve the Fermi-arc splitting\cite{rao2019observation,sanchez2019topological}. Thus, within the experimental tolerance, the electronic states near the Fermi level are well described by a model without SOC. 

\textcolor{black}{Lastly, we will comment on existing studies on materials with well-documented grain boundary characterization which are now known to be topological in nature. A prime example is Antimony (Sb), which has been identified as a topological semimetal\cite{zhang2012topological,bian2012interfacial,cairns2015observation}. Several studies have characterized Sb thin films as polycrystalline with smaller grains found in thinner films\cite{deschacht1980effect,de1983electric}. If materials such as Sb were to be used as interconnects with very small dimensions (on the order of nanometers), the grain size would be expected to be quite small. Our results show that grain boundary resistivity decreases for these very thin films. Thus, grain boundaries would be expected to become more prevalent but less resistive in highly-scaled topological semimetals such as Sb. Since these trends are in opposite directions, the scaling of total resistivity would ultimately depend on the balance between the two and the details of the surface transport.}

\section{Conclusion}
We have investigated the electron scattering characteristics at grain boundaries in thin films of two topological semimetals (CoSi and CoGe) evaluated as potential materials for next-generation interconnects. We find that relative to conventional interconnect metals like Cu and Al, the bulk grain boundary specific resistivity of CoSi and CoGe is roughly two orders-of-magnitude larger due to the significantly lower bulk DOS. On the other hand, in CoSi and CoGe thin films with thickness up to 5nm, we find that the grain boundary specific resistivity decreases with decreasing film thickness, opposite to the trend in conventional metals Cu and Al. This is because the electron transport in the topological semimetals CoSi and CoGe is primarily surface-dominated for both pristine structure and those containing grain boundaries. This different size scaling behavior of grain boundary specific resistivity in the topological semimetals sets them apart from conventional metals in terms of the resistivity bottleneck at critically small dimensions, thus suggesting that they may hold promise in advanced interconnect technologies where grain boundaries are prevalent and resistivity reduction is essential for high-performance. It is hence imperative that we explore a wide variety of topological semimetals to identify materials with a sufficiently large density of topological surface states, so that not only the scaling trend, but also the magnitude of the grain boundary specific resistivity can drop below that of conventional interconnect metals like Cu for highly-scaled technology applications. 

\textcolor{black}{Future experimental studies will focus on investigating grain boundary density in CoSi thin films and the impact on resistivity scaling. Future theoretical studies should focus on identifying topological materials with a maximum number of surface conducting channels. This could include materials with maximum number of pairs of Weyl nodes, maximum Chern number, maximum Fermi arc length in the Brilluoin Zone or maximum density of Fermi arcs in the Brillouin Zone. }

%\bibliography{topological} 


\end{document}